\documentclass[paper]{geophysics}
\usepackage{chemformula}
\usepackage{hyperref}
\usepackage{amsmath}
\usepackage{amssymb}
\usepackage{amsfonts}

\begin{document}
\title{DeepNRMS: Unsupervised Deep Learning for Noise-Robust \ch{CO2} Monitoring in Time-Lapse Seismic Images}

\renewcommand{\thefootnote}{\fnsymbol{footnote}} 

\address{Stanford University}
\author{Min Jun Park, Julio Frigerio, Bob Clapp, and Biondo Biondi\\Stanford University}
% 

% \righthead{}

\maketitle
\begin{abstract}
Monitoring stored \ch{CO2} in carbon capture and storage projects is crucial for ensuring safety and effectiveness. We introduce DeepNRMS, a novel noise-robust method that effectively handles time-lapse noise in seismic images. The DeepNRMS leverages unsupervised deep learning to acquire knowledge of time-lapse noise characteristics from pre-injection surveys. By utilizing this learned knowledge, our approach accurately discerns \ch{CO2}-induced subtle signals from the high-amplitude time-lapse noise, ensuring fidelity in monitoring while reducing costs by enabling sparse acquisition. We evaluate our method using synthetic data and field data acquired in the Aquistore project. In the synthetic experiments, we simulate time-lapse noise by incorporating random near-surface effects in the elastic properties of the subsurface model. We train our neural networks exclusively on pre-injection seismic images and subsequently predict \ch{CO2} locations from post-injection seismic images. In the field data analysis from Aquistore, the images from pre-injection surveys are utilized to train the neural networks with the characteristics of time-lapse noise, followed by identifying \ch{CO2} plumes within two post-injection surveys. The outcomes demonstrate the improved accuracy achieved by the DeepNRMS, effectively addressing the strong time-lapse noise. 
\end{abstract}

%--------------------------------------------------------------------%

\section{Introduction}
To mitigate the worst consequences of climate change, achieving net-zero global greenhouse gas emissions by the mid-century is of utmost importance \cite[]{IPCCPolicymakers2018}. Carbon capture, utilization, and storage (CCUS) have emerged as essential strategies for attaining this objective \cite[]{Agency2019}. Among the various CCUS approaches, injecting carbon dioxide (\ch{CO2}) into subsurface reservoirs for permanent storage holds great promise as a large-scale solution for reducing \ch{CO2} emissions \cite[]{firoozabadi2010prospects, krevor2023subsurface}. However, ensuring the safety and efficiency of \ch{CO2} storage projects necessitates accurate long-term monitoring and verification systems. Time-lapse seismic monitoring, also known as 4D seismic, is a powerful method for achieving these objectives \cite[]{lumley20104d}. This method involves the repeated acquisition of seismic data at given intervals over a period of time. One major benefit is its ability to provide a detailed, high-resolution view of the subsurface, enabling the detection of small changes in its elastic properties. Numerous studies have demonstrated the ability of time-lapse seismic monitoring to detect \ch{CO2}-induced changes in the subsurface \cite[]{arts2003monitoring, chadwick2010quantitative,urosevic2011seismic, bacci2017using, roach2017initial, white20197}. In seismic data, the signal of these subsurface changes typically manifests as subtle, low-amplitude signals. Furthermore, the data often contain other signals, considered to be noise. Between seismic surveys, other subsurface changes unrelated to the \ch{CO2} injection almost invariably occur, such as alterations in the near-surface weathering zone. Another source of time-lapse noise is the footprints arising from non-repeatability, stemming from different acquisition parameters across surveys. Successful monitoring can only be achieved if the time-lapse signal induced by the injected \ch{CO2} can be distinguished from the noise generated by these and other factors.

Hence, improving repeatability is important to distinguish the subtle time-lapse signal from the time-lapse noise \cite[]{lumley20104d}. However, it generally requires a higher cost for both acquisition and processing, limiting the capability of long-term continuous monitoring \cite[]{houck2007time,ringrose2013salah}. Although sparse acquisition has been suggested as a solution to this issue \cite[]{white2015time,roach2015assessment}, attaining high fidelity without high-density source and receiver acquisition can be challenging \cite[]{bakulin2016processing, smith20184d}. Consequently, effectively handling time-lapse noise becomes crucial to achieving both high fidelity and cost-effectiveness in monitoring endeavors. 

Recent advancements in computational power have significantly facilitated the application of deep learning in various geophysical domains, including lithology prediction \cite[]{zhang2018deep, zhang2020seismic}, seismic data processing \cite[]{wang2019deep, yu2019deep, matharu2020simultaneous, park2020seismic}, velocity analysis \cite[]{araya2018deep, park2020automatic, fabien2020seismic}, and enhancing repeatability in time-lapse data \cite[]{alali2022time, jun2022repeatability}. Deep neural networks (DNN) have demonstrated their efficacy in extracting meaningful features directly from high-dimensional geophysical data. Recently, deep learning approaches have also been proposed for monitoring \ch{CO2} plumes \cite[]{zhou2019data, yuan2020time, sheng2022deep, biondi2022integration}. \cite{zhou2019data} employed a deep supervised learning approach to predict \ch{CO2} leakage mass using seismic data, while \cite{yuan2020time} predicted velocity changes using time-lapse seismic data. However, generating realistic training data, including realistic noise and scenarios that accurately represent \ch{CO2} plume behaviors for real data applications, remains a challenging task. It is known as the generalization problem in deep learning studies for geophysical problems \cite[]{li2019deep, yu2021deep, wu2021deep,jun2022realistic}. To address this challenge, \cite{sheng2022deep} proposed a method to incorporate real noise into synthetic training data, which improved the alignment of the trained model's predictions with manually interpreted \ch{CO2} plumes in field data. However, their approach relies on intensive workflows to generate synthetic training data and primarily focuses on data with a relatively strong signal-to-noise ratio. Therefore, further enhancements and validations are required to tackle the challenges posed by strong time-lapse noise effectively.

In this paper, we present a novel unsupervised approach that utilizes an anomaly detection scheme. We define time-lapse noise as the \textbf{"normal data"} and the time-lapse signals (i.e., \ch{CO2} plume signals) as the \textbf{"anomaly data."} Specifically, we propose DeepNRMS, an unsupervised deep learning method designed to accurately localize the \ch{CO2} plume amidst strong time-lapse noise. Our approach aims to leverage the distinctive features of time-lapse noise extracted from pre-injection surveys to enhance the identification of the time-lapse signal associated with stored \ch{CO2}. By effectively separating the \ch{CO2} response from time-lapse noise, our approach has the potential to improve monitoring accuracy and reduce monitoring costs for noisy data. To validate the effectiveness of our proposed approach, we conduct extensive evaluations using both 2D synthetic data and 3D field data obtained from the Aquistore project \cite[]{white2015time, roach2015assessment, roach2017initial}. It is crucial to note that our method for Aquistore data works upon the availability of two pre-injection surveys. The results demonstrate that incorporating learned knowledge from the pre-injection surveys significantly enhances the performance of \ch{CO2} plume monitoring, even in the presence of time-lapse noise.

%--------------------------------------------------------------------%

\section{Methods}
The DeepNRMS is a new method that combines the normalized root mean squared difference (NRMS) metric with an anomaly score acting as a weighting function. The NRMS metric quantifies the dissimilarity between pre-injection and post-injection seismic images, providing a baseline assessment of \ch{CO2}-induced changes \cite[]{kragh2002seismic}. NRMS difference between two different images ($d_1$ and $d_2$) can be calculated as follows: 

\begin{equation}
\mathrm{NRMS(d_1,d_2)}=\frac{2 \times \operatorname{\sqrt{\frac{\sum(d_1-d_2)^2}{\mathrm{~N}}}}}{\operatorname{\sqrt{\frac{\sum d_1^2}{\mathrm{~N}}}}+\operatorname{\sqrt{\frac{\sum d_2^2}{\mathrm{~N}}}}},
\end{equation}

where $N$ is the total number of samples. While NRMS can offer a reliable location of the injected \ch{CO2} plume, the presence of strong time-lapse noise can impede accurate \ch{CO2} plume detection using NRMS alone. To address this issue, the DeepNRMS incorporates NRMS with an anomaly score, which can identify regions in the seismic image that deviate significantly from normal patterns. To obtain the anomaly score, we adopt the Deep-Support Vector Data Description (Deep-SVDD) algorithm \cite[]{ruff2018deep}, an unsupervised anomaly detection approach. By filtering the NRMS using the obtained anomaly score, the DeepNRMS enhances the visibility and clarity of \ch{CO2} plumes, improving detection and monitoring accuracy. In other words, the integration of NRMS and the anomaly score enables comprehensive analysis of time-lapse seismic images, effectively suppressing noise and emphasizing \ch{CO2}-induced signals.

\subsection{Deep-SVDD}
\begin{figure}[ht!]
\centerline{\includegraphics[width=1\textwidth]{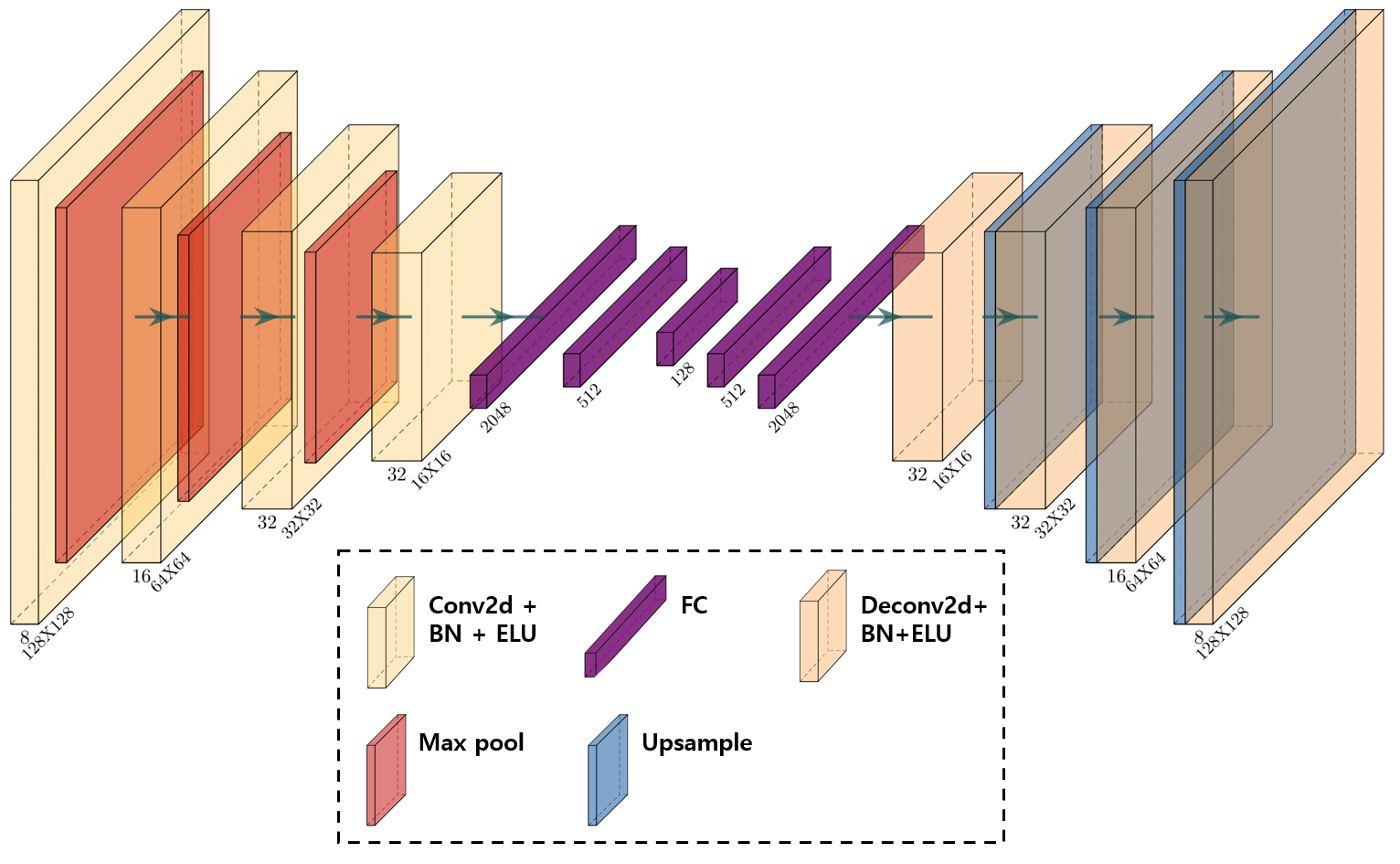}}
\caption{The architecture of the autoencoder we adopt. The encoder on the left carries out the feature extraction process, while an attempt is made to reconstruct the input by the decoder on the right.}
\label{fig:autoencoder}
\end{figure}

Deep-SVDD is an unsupervised deep learning approach designed for anomaly detection. Its primary function is to discern patterns or instances within a dataset that diverge from the anticipated or common behavior \cite[]{ruff2018deep}. For the implementation of Deep-SVDD in our study, we begin by utilizing normal data to train a convolutional autoencoder, denoted as \(\phi_{DNN}\). This procedure aims to understand the compact embedding representations, depicted as 128 vectors in Figure \ref{fig:autoencoder}, derived from complex input data. 

The architecture of \(\phi_{DNN}\) can be observed in Figure \ref{fig:autoencoder}, comprising both an encoder and a decoder. The encoder is structured with four convolutional blocks, each accompanied by three fully connected (FC) layers. Every convolutional block is made up of a 2D 3\(\times\)3 convolutional kernel, integrated with batch normalization (BN) and an exponential linear unit (ELU). Following this, a max pooling layer is added \cite[]{clevert2015fast}. On the other hand, the decoder is composed of two FC layers and an additional four deconvolutional blocks. These blocks are crucial for reconstructing the original data dimensions derived from the embedding representations. Each of these deconvolutional blocks integrates a 2D 3\(\times\)3 transposed convolution layer, accompanied by BN, ELU, and an upsampling layer. The training process for \(\phi_{DNN}\) is to get the optimal weights ($\hat{\theta}$), which is represented as follow:

\begin{equation}
\hat{\theta}=\arg \min _{\theta} \sum_{i=1}^N L\left(x_i, \phi_{DNN}\left(x_i ; \theta\right)\right),
\label{eq:argmin}
\end{equation}
where $N$ is the total number of data. We define a loss function, $L$, as a mean squared error (MSE) defined as the following equation:

\begin{equation}
\mathrm{L_{MSE}}=\frac{1}{N} \sum_{i=1}^N\left(x_i-\hat{x}_i\right)^2.
\label{eq:mse}
\end{equation}

Once the model has been trained, we can use a trained encoder to extract the 128 vectors that are the embedding representations from the original input data. The second step is to get the sphere's center ($\boldsymbol{c}$) by averaging those embedding representations of the normal training data. Then, with the same normal data as input, we update the trained encoder by minimizing the following objective function:

\begin{equation}
 \min_{\theta}. \frac{1}{n} \sum_{i=1}^{n} \left\|\phi_{ENC}\left(\boldsymbol{x}_i ; \theta\right)-\boldsymbol{c}\right\|^2 + \frac{\lambda}{2}||\theta||^2,
\label{eq:deep-svdd}
\end{equation}

where $\lambda$ is a weighting parameter that controls the strength of the regularization term, $\theta$ is the weights of the encoder of pre-trained DNNs ( $\phi_{ENC}$). The goal is to get the optimal encoder to minimize the distance between the output of the encoder, represented by $\phi_{ENC}\left(\boldsymbol{x}_i ; \theta\right)$, and the center of the sphere, $\boldsymbol{c}$. After determining the optimal weights ($\theta^*$), we predict the anomaly score of given new instances (test data) by calculating the distance from the center, $\left\|\phi_{ENC}\left(\boldsymbol{x}; \theta^*\right)-\boldsymbol{c}\right\|^2$. Finally, we derive the DeepNRMS by multiplying the NRMS metric with the anomaly scores.

%--------------------------------------------------------------------%
%--------------------------------------------------------------------%

\section{Synthetic example}

\subsection{Workflows}
\begin{figure*}[h]
\centerline{\includegraphics[width=0.7\textwidth]{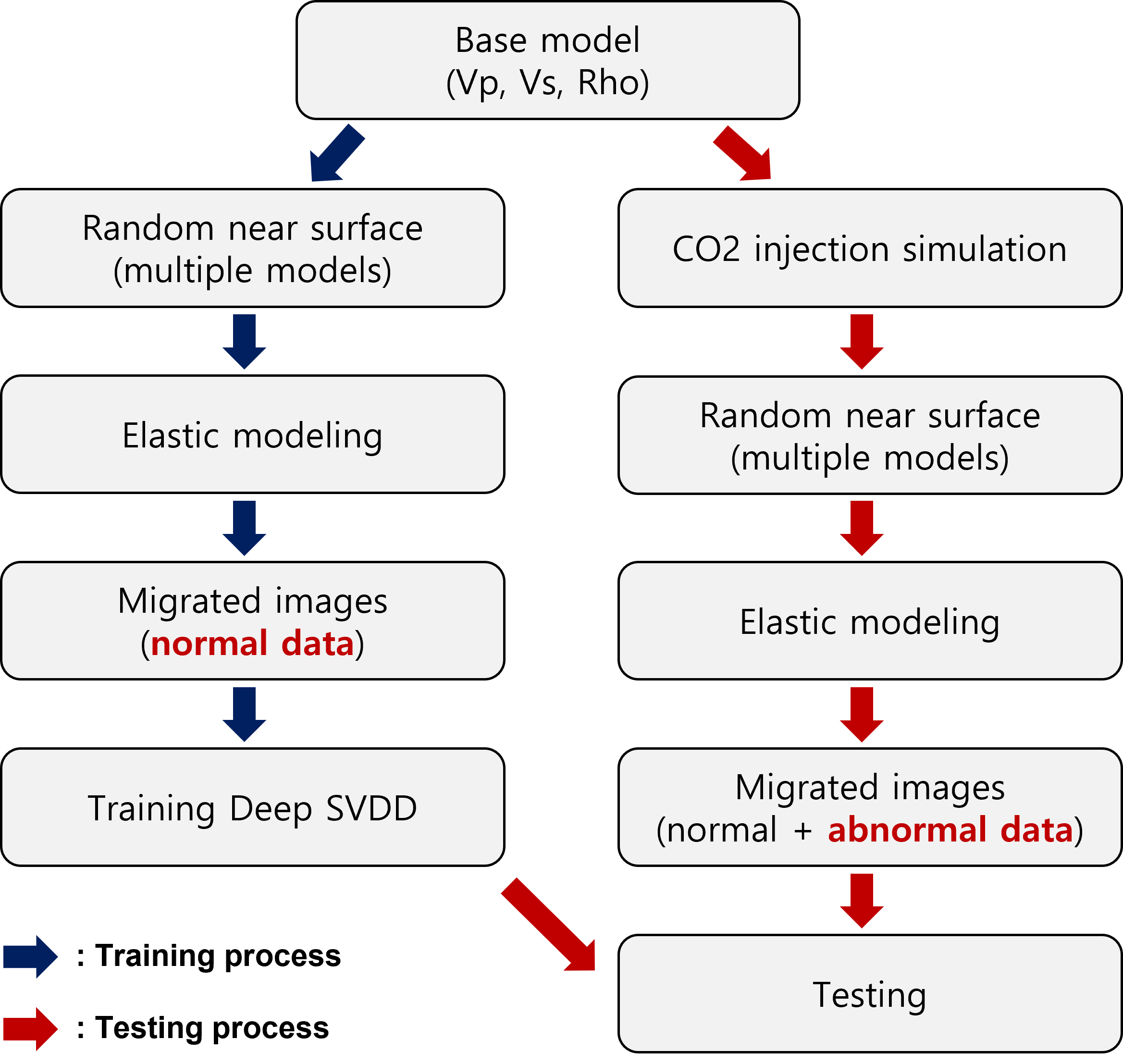}}
\caption{Proposed workflows for synthetic examples include two branches: training (blue arrows) and testing (red arrows). The training process only requires normal data (no injected \ch{CO2}), which can be generated by giving perturbations at the near-surface.}
\label{fig:wf}
\end{figure*}

Our method employs a two-branch workflow, encompassing both training and testing processes, as depicted in Figure \ref{fig:wf}. In the training phase, we simulate realistic time-lapse noise by introducing near-surface perturbations to the base model. Next, we generate seismic data through elastic modeling and migrate the data to obtain seismic images, which serve as our training data (referred to as normal data). It is important to note that the training data are obtained by differencing the baseline image ($i^{0}$) with other pre-injection images.

The testing phase follows a similar procedure to the training step, with the exception that we simulate \ch{CO2} injection prior to introducing time-lapse noise. Consequently, the images generated in the testing branch include abnormal data, reflecting the effects of injected CO2. Similar to the training data, the testing data are derived by differencing $i^{0}$ with the post-injection images. Our workflow aims to detect anomalies induced by injected \ch{CO2} using Deep-SVDD, which is trained solely on the pre-injection images. The resulting anomaly scores are then utilized as filters for the NRMS, leading to the development of the DeepNRMS.

\subsection{Training}
\begin{figure}[ht!]
\centerline{\includegraphics[width=1\textwidth]{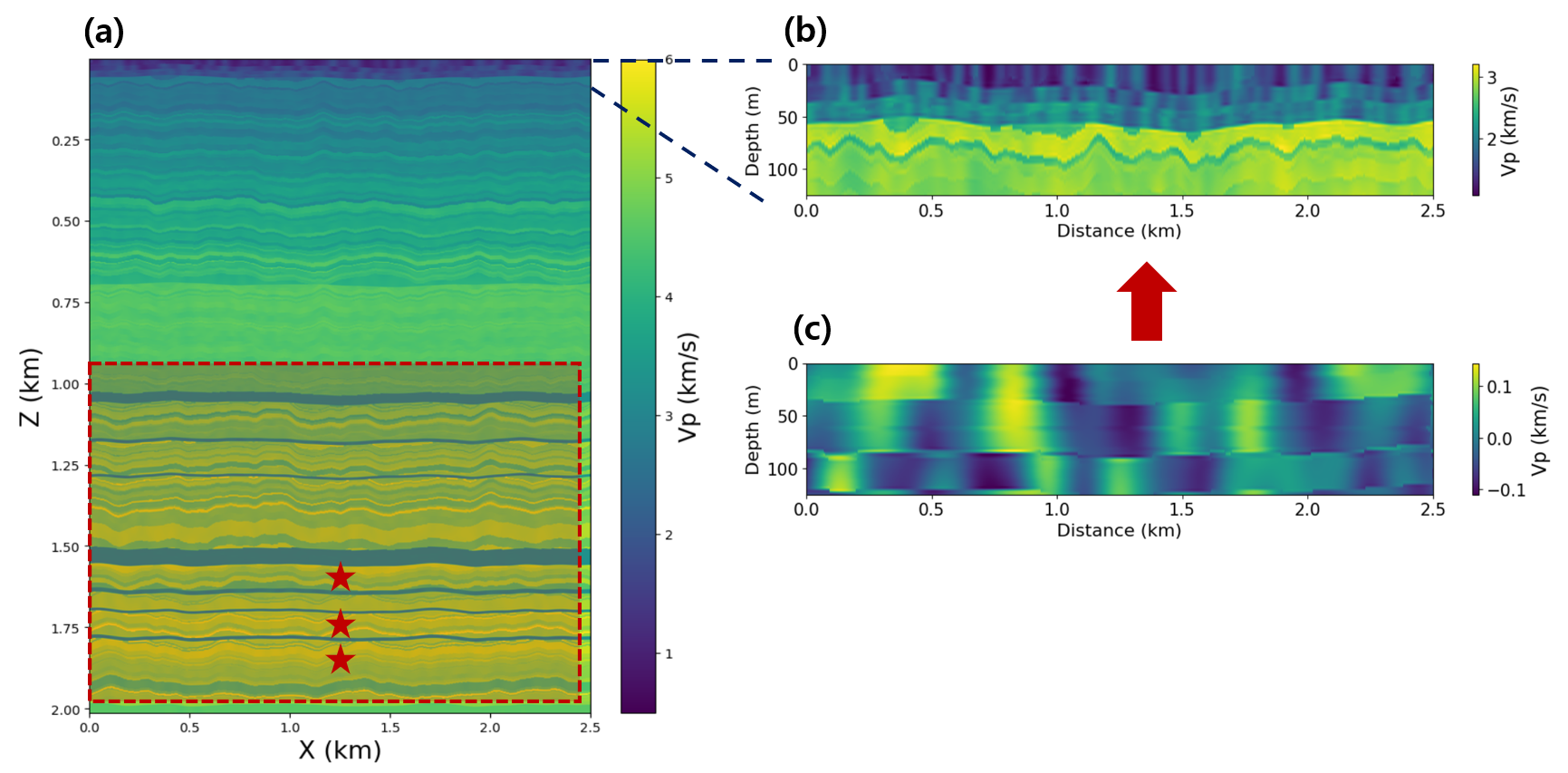}}
\caption{The \ch{CO2} storage model. (a) The $v_p$ model after \ch{CO2} injection. Red stars indicate three different injection locations, and the red dashed area shows the monitoring area. (b) The zoomed-in section of the near-surface. (c) One of the random Vp perturbation (-100$m/s$ to +100$m/s$). Every survey has its unique near-surface properties. 
}
\label{fig:model}
\end{figure}

As discussed earlier, the training data in this example is taken only from the \ch{CO2} pre-injection images. To get realistic time-lapse noises, we simulate the low-velocity zone near the surface and give random perturbations for each survey. Figures \ref{fig:model} show (a) the P-velocity model and (b) the zoomed-in section of the near-surface with (c) its perturbation ranging from -100$m/s$ to +100$m/s$. Red stars in Figure \ref{fig:model} (a) indicate three different injection locations. To save computational cost, we select a certain part of the model (red dashed area) for the monitoring area. We then simulate the seismic data using elastic wave modeling and produce the images by performing Kirchhoff prestack depth migration. 

\begin{table}[h!]
\centering
\begin{tabular}{ |p{3cm}|p{3cm}||p{3cm}|p{3cm}|  }
\hline
Name & Parameter & Name & Parameter \\
\hline
\hline
Shot number   & 25 & Total time step & 1500   \\
Shot spacing & 100 $\mathrm{m}$& Time interval & 0.001 sec \\
Receiver number & 500 & Model spacing & 2.5 $\mathrm{m}$\\
Receiver spacing & 5 $\mathrm{m}$& Wavelet & Ricker 25 $\mathrm{Hz}$ \\
\hline
\end{tabular}
\caption{The parameters for both the synthetic modeling and migration.}
\label{table:1}
\end{table}

Table \ref{table:1} shows the parameters for both modeling and migration. We used a 25 $\mathrm{Hz}$ Ricker wavelet for modeling. For each survey, we used a total of 25 shots with 100 $\mathrm{m}$ spacing. Receiver spacing is 5 $\mathrm{m}$, and the time interval is 0.001 sec. Once pre-injection images are generated, we perform the repeatability analysis using the NRMS difference. For the repeatability analysis, we define the baseline image ($i^{0}$) and calculate the NRMS difference between $i^{0}$ and other pre-injection images with a 10 $\mathrm{m}$ window size. 

\begin{figure}[ht!]
\centerline{\includegraphics[width=1\textwidth]{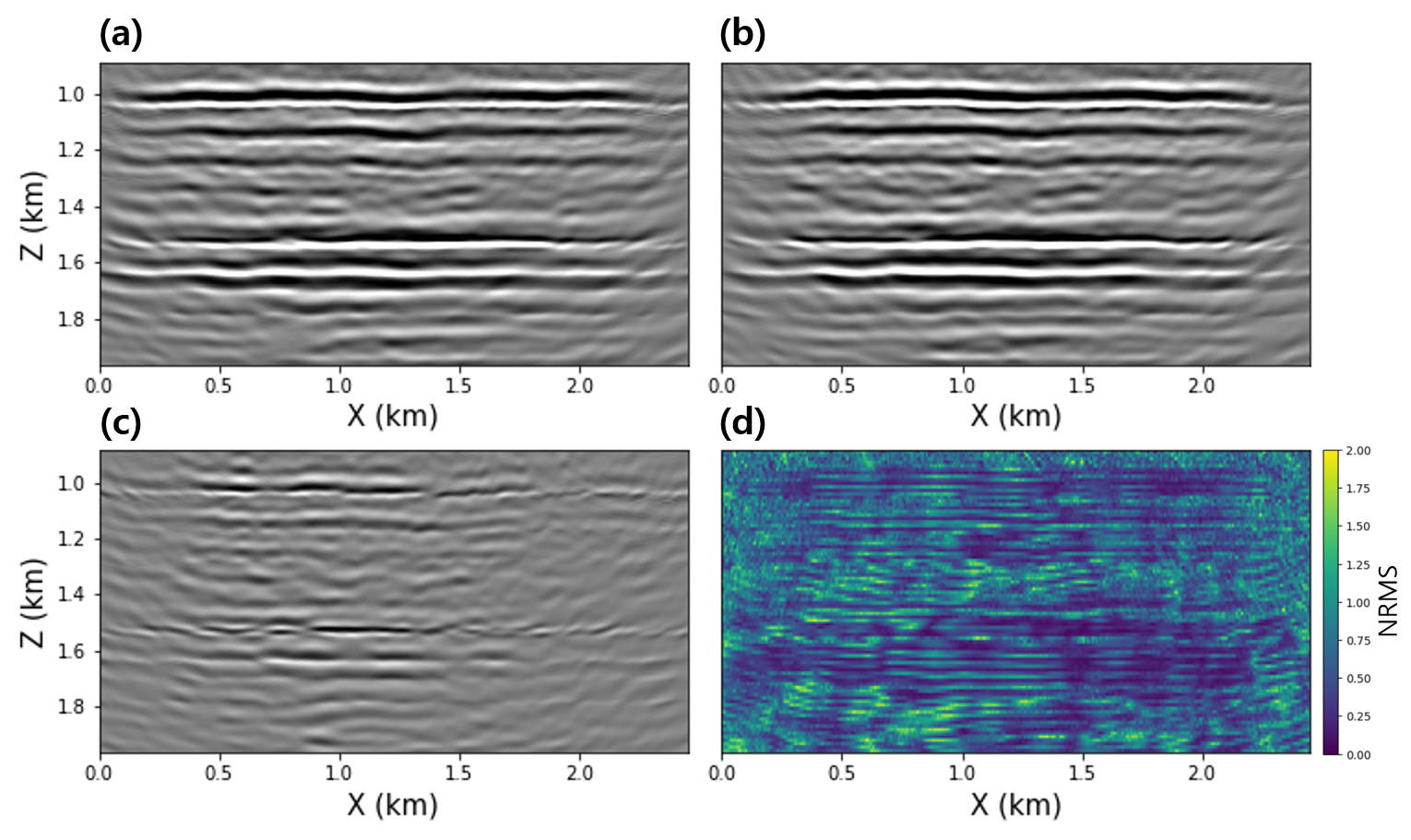}}
\caption{(a) The baseline image ($i^{0}$), (b) another pre-injection image, (c) their amplitude difference, and (d) their NRMS difference are shown.}
\label{fig:image_pre}
\end{figure}

Figure \ref{fig:image_pre} shows an example of the analysis result showing (a) the baseline image, (b) another pre-injection image, (c) their amplitude difference, and (d) their NRMS difference. Despite the pre-injection stage, we can see strong time-lapse differences caused by the near-surface perturbation. In general, NRMS values lower than $0.2$ are considered excellent repeatability \cite[]{lumley20104d}. The average NRMS value of the total pre-injection dataset is 0.46, showing relatively higher time-lapse noise. Finally, we extract the $128$ by $128$ size ($320$ $\mathrm{m}$ by $320$ $\mathrm{m}$) of patches from the amplitude difference between $i^{0}$ and $100$ of pre-injection images for training the Deep-SVDD. 

In the training step, we first train the autoencoder (Figure \ref{fig:autoencoder}) by minimizing Equation \ref{eq:argmin}. The learning rate is $0.0001$, and the total number of epochs is $100$. Secondly, we use the trained encoder to project the $128$ size of the latent vector from the patches and calculate the sphere's center ($\boldsymbol{c}$) by averaging those vectors. Lastly, we fine-tuned the encoder by solving Equation \ref{eq:deep-svdd}. We adopt $0.0001$ for the learning rate for fine-tuning, and the number of epochs is $20$.

\subsection{Testing}
\begin{figure}[ht!]
\centerline{\includegraphics[width=1\textwidth]{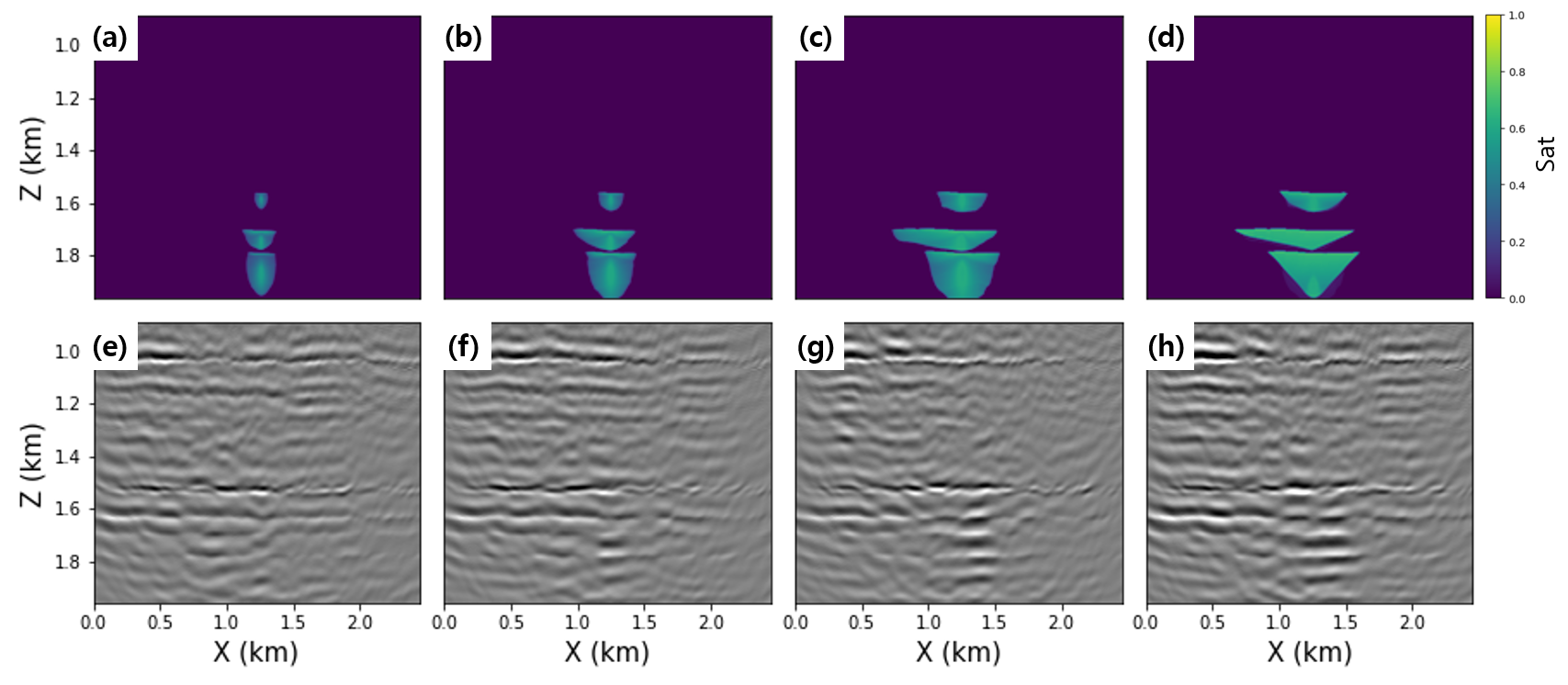}}
\caption{The first row (a,b,c,d) shows the snapshots of \ch{CO2} saturation maps resulting from the GEOS simulator. The second row (e,f,g,h) shows the amplitude differences between the baseline image,$i^{0}$, and $4$ of post-injected images.}
\label{fig:co2sat}
\end{figure}

To test the trained Deep-SVDD, we first simulate the \ch{CO2} injection using GEOS (\url{http://www.geosx.org/}), an open-source multiphysics simulator. Figures \ref{fig:co2sat} (a),(b),(c), and (d) show the snapshots of \ch{CO2} saturation maps resulting from the GEOS simulator. We update the base models using the four distinct \ch{CO2} saturations, applying the Gassmann equation to estimate changes when \ch{CO2} substitutes water in the pore space \cite[]{Park.sep.184.minjun1}. Then, we generate the images by performing the migration using the same parameters of the pre-injection imaging. Finally, we get the anomaly score ($\left\|\phi_{ENC}\left(\boldsymbol{x}; \theta^*\right)-\boldsymbol{c}\right\|^2$) of the patches extracted from those amplitude differences between the baseline image,$i^{0}$, and $4$ of post-injection images (Figures \ref{fig:co2sat} (e),(f),(g), and (h)).

\begin{figure}[h!]
\centerline{\includegraphics[width=1\textwidth]{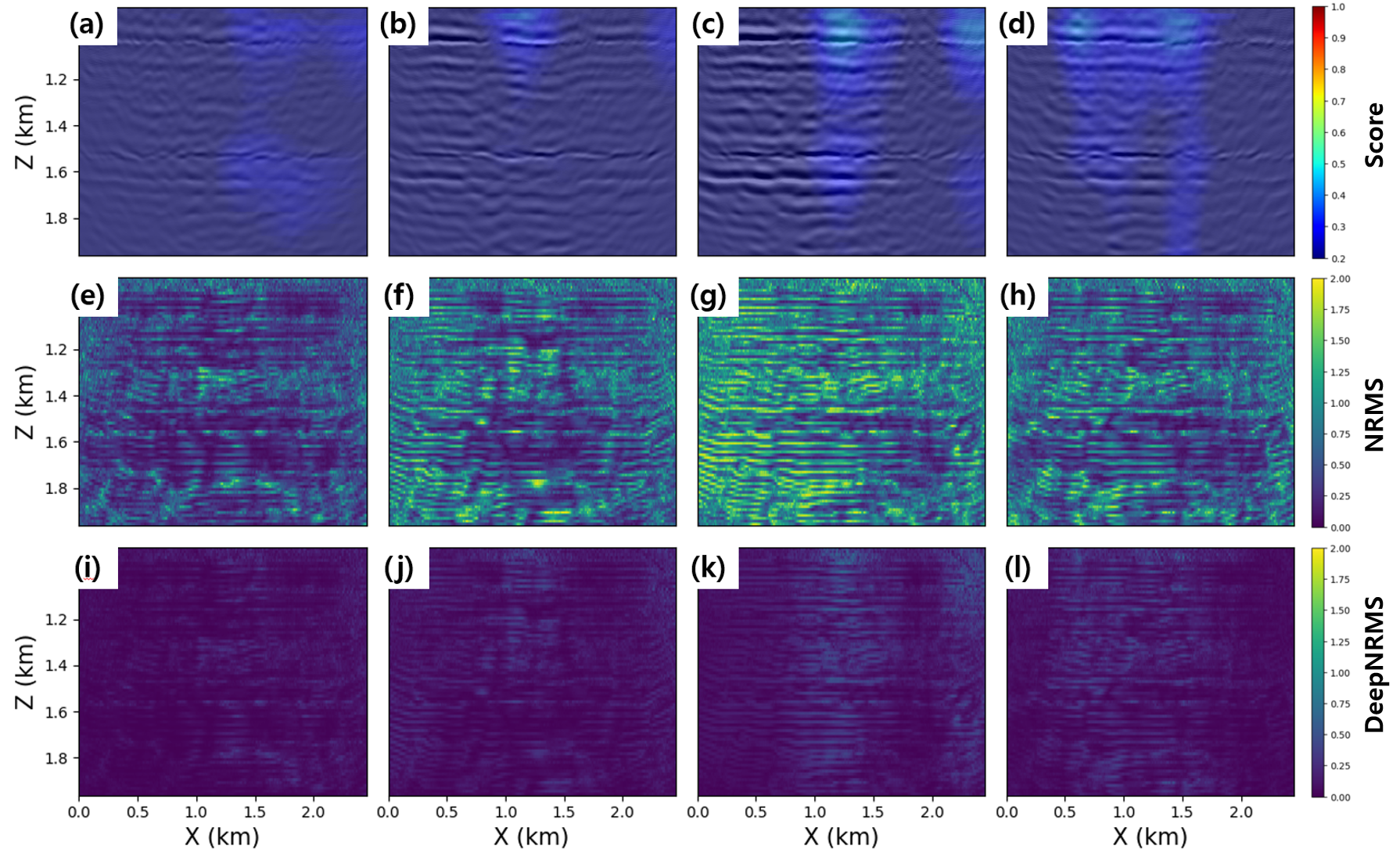}}
\caption{The prediction results on the four pre-injection images are shown.
The top row (a,b,c,d) shows the amplitude differences superposed by the anomaly scores. The second row (e,f,g,h) indicates the NRMS differences between the baseline image and each monitoring image. The third row (i,j,k,l) shows the DeepNRMS differences. Note that the pre-injection images in the figures were not used for the training.}
\label{fig:pred_pre}
\end{figure}

First, we test the trained model with the pre-injection images that were not used for the training. Figures \ref{fig:pred_pre} (a,b,c,d) shows the four different pre-injection images with the anomaly scores. As we can see, the trained model gives a lower score in most regions even though high amplitude differences exist. Thus, the NRMS differences multiplied by the anomaly scores (Figure \ref{fig:pred_pre} (i,j,k,l)) show smaller NRMS differences values compared to the original NRMS differences (Figure \ref{fig:pred_pre} (e,f,g,h)). These results imply that learning the characteristics of the time-lapse noise can help to identify the \ch{CO2} plumes that have not existed in the training data.  

\begin{figure}[h!]
\centerline{\includegraphics[width=1\textwidth]{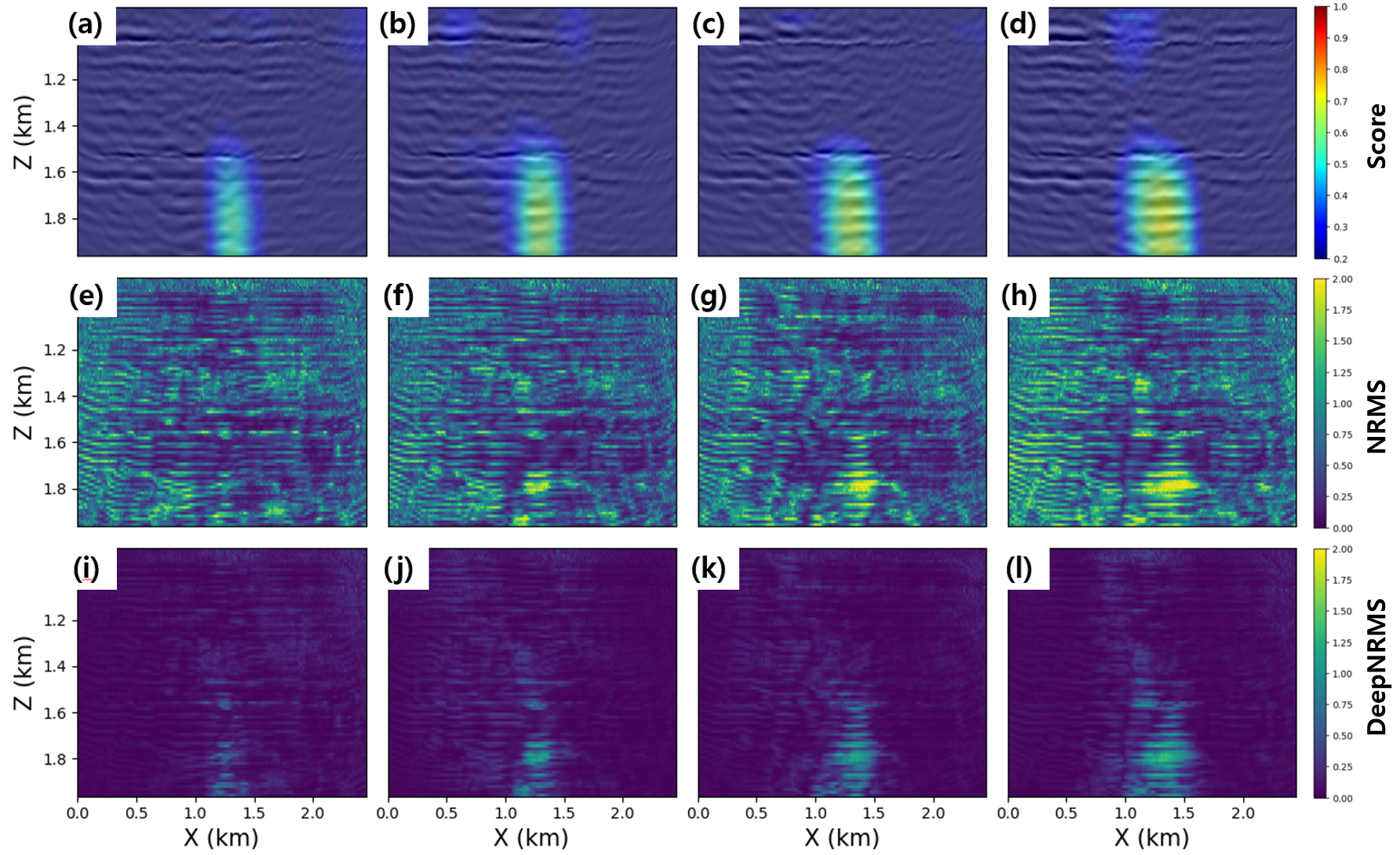}}
\caption{The prediction results on the four post-injection images are shown. The figures' structures are the same as Figure \ref{fig:pred_pre}.}
\label{fig:pred_post}
\end{figure}

Next, we test the same trained model with the four post-injection images. Figures \ref{fig:pred_post} show the prediction results on the four post-injection images (Figure \ref{fig:co2sat} (e),(f),(g), and (h)). The top row (a,b,c,d) shows the amplitude differences superposed by the anomaly score. As we can see, the trained model provides the higher anomaly score where the plumes are located while giving lower scores for others. The second row (e,f,g,h) indicates the NRMS differences between the baseline image and each monitoring image. The third row (i,j,k,l) shows the DeepNRMS differences. We can see that the DeepNRMS successfully remove most of the strong time-lapse noise (or NRMS difference) while boosting the signal caused by the stored \ch{CO2} plumes.

%--------------------------------------------------------------------%
\section{Field example}
\subsection{Aquistore \ch{CO2} storage site}
Aquistore is a \ch{CO2} storage site in the vicinity of the town of Estevan, in the province of Saskatchewan, Canada. The \ch{CO2} is captured from the coal-fired SaskPower Boundary Dam power plant \cite[]{roach2015assessment} and injected into a saline aquifer at depths of $3150$-$3350$ $\mathrm{m}$(1.8-1.9 sec). The reservoir is composed of sandstones, with silty-to-shaley and carbonate interbed, of the Deadwood and Winnipeg Formations, the latter overlying the former. At the top, it is sealed by a layer of shales of the Winnipeg Formation. The injection of \ch{CO2} started in the Fall of 2015 and was sustained at rates of $400$-$600$ tonnes/day. At the beginning of 2019, the total amount of \ch{CO2} injected was approximately $200$ kilotonnes \cite[]{white20197}. Seismic monitoring of the \ch{CO2} injection and storage into the reservoir was conducted through time-lapse seismic with six surface-seismic surveys being acquired. To reduce costs, the receiver distribution was configured sparsely. In this context, $\boldsymbol{sparse}$ denotes a reduced number of geophones and shots deployed per unit area over the survey relative to the deployment density in a state-of-the-art 3D seismic survey \cite[]{roach2017initial}. Moreover, the geophone receivers were permanently buried at $20$ $\mathrm{m}$ depth to increase repeatability between surveys, and the vertical component geophones were recorded.  

\begin{table}[h!]
\centering
\begin{tabular}[c]{|p{2cm}|p{2cm}|p{2cm}|p{5cm}|}
\hline
Survey & Date & \ch{CO2} (kT) & Noise mean $\pm$ std dev (dB)\\
\hline
\hline
PI1 &  Mar/2012 & Zero & $-28.19 \pm 2.28$ \\
PI2 &  Nov/2013 & Zero & $-28.74 \pm 2.11$ \\
M1 &  Nov/2016 & 102  & $-27.81 \pm 2.22$ \\
M2 &  Mar/2018 & 141  & $-27.54 \pm 1.96$ \\
\hline
\end{tabular}
\caption{Surface seismic surveys dates and amount of \ch{CO2} injected until the occasion of the acquisition (Adapted from \cite{white20197}).}
\label{table:seismic_dates_AmountCO2}.
\end{table}

Table \ref{table:seismic_dates_AmountCO2} shows the dates, the amount of injected \ch{CO2}, and the average RMS ambient noise levels of the surveys, including two pre-injection surveys (PI1 and PI2) and two post-injection surveys (M1 and M2). For those surveys, the receiver locations were the same, and the source locations had an uncertainty of 1-2 $\mathrm{m}$. The source type was dynamite with a charge size of 1 kg. The area covered by the sources and receivers was \(3.0 \times 3.0\) and \(2.5 \times 2.5\) $\mathrm{km}$, respectively. The space between the shot and receiver lines was 288 $\mathrm{m}$ and 144 $\mathrm{m}$, while the inline distance between shots and receivers was 144 $\mathrm{m}$ and 72 $\mathrm{m}$, respectively. Each dataset received an independent but identical pre-stack processing and post-stack 3D time migration \cite[]{roach2017initial}. Following the migration, each monitor image was cross-equalized with the baseline image using phase-time matching, shape filtering, and amplitude normalization. Finally, time shifts were applied to match the time of the topo of the reservoir between each monitor and baseline.

\begin{figure}[hb!]
\centerline{\includegraphics[width=1\textwidth]{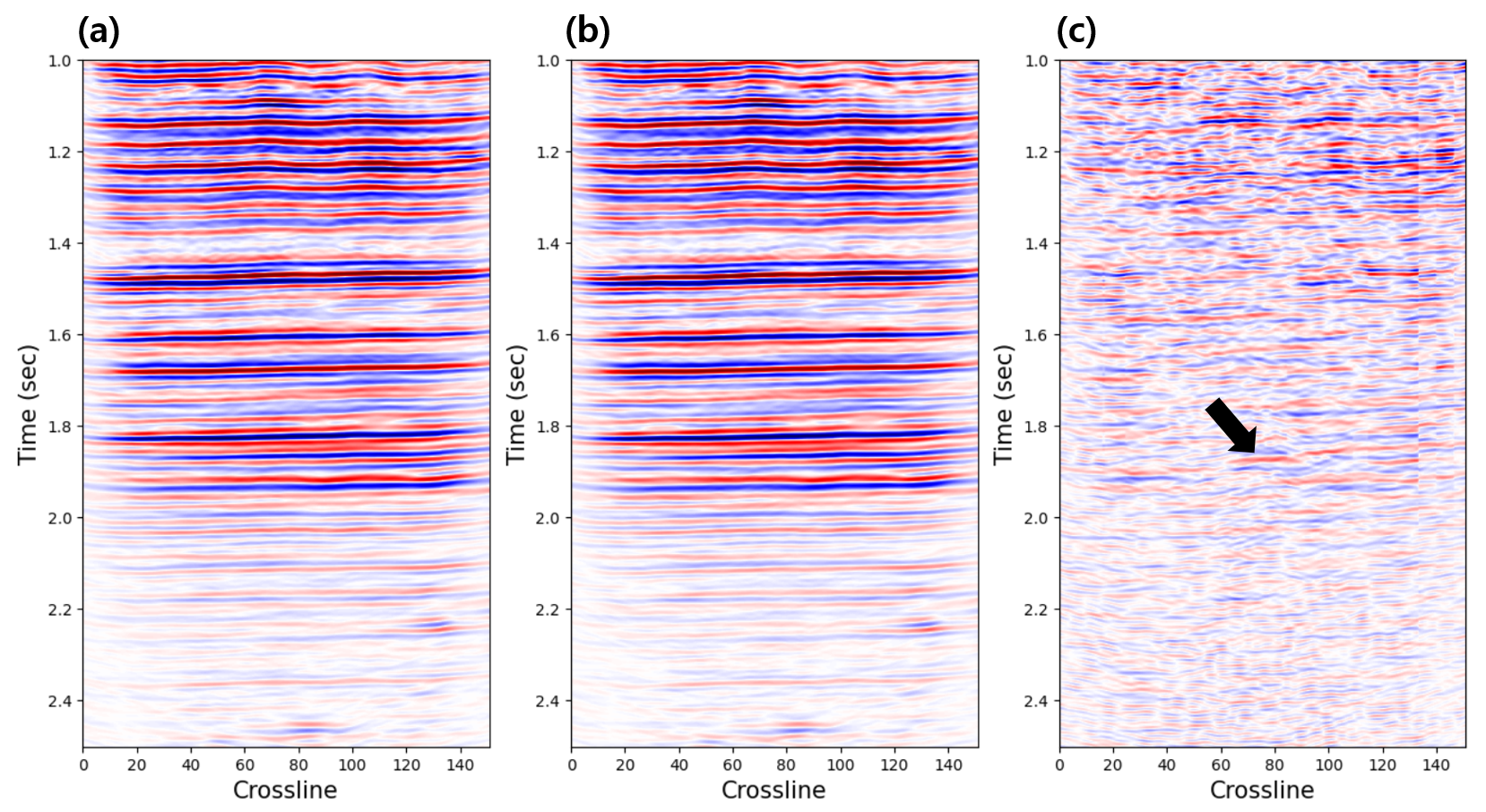}}
\caption{The inline section 75 of (a) M2, (b) PI1, and (c) their amplitude differences are shown. The black arrow indicates the signal from the injected \ch{CO2} plume. Note that the amplitude difference (c) is amplified by a factor of five for visualization.}
\label{fig:aquistore1}
\end{figure}

Figures \ref{fig:aquistore1} show the inline section of the migrated image of (a) M2, (b) PI1, and (c) their difference. The black arrow indicates the injected \ch{CO2} plume. Note that the amplitude difference of Figure \ref{fig:aquistore1} (c) is amplified by a factor of five for visualization. Despite careful processing and high enough repeatability of the data acquisition, time-lapse noises are still comparable to the plume signal. This section aims to learn the characteristics of the time-lapse noise from the two pre-injection surveys (PI1 and PI2) and then use that knowledge to remove the time-lapse noise from the post-injection surveys (M1 and M2) in order to determine the position of the plume more clearly.

\subsection{Results}
\begin{figure}[hb!]
\centerline{\includegraphics[width=1\textwidth]{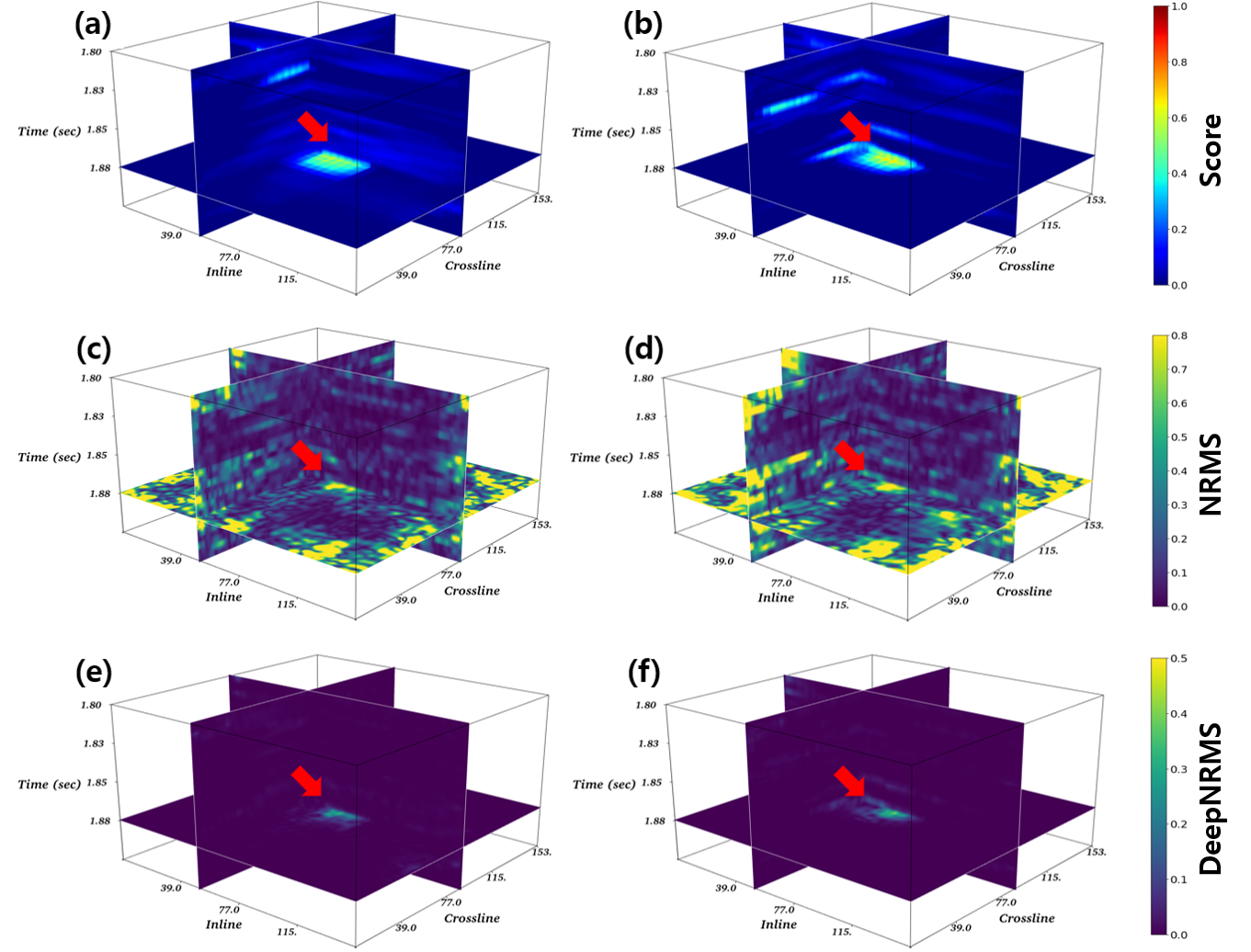}}
\caption{The top row shows the anomaly scores for (a) M1 and (b) M2. The second row indicates the NRMS differences for (a) M1 and (b) M2. The bottom row shows the DeepNRMS for (c) M1 and (d) M2.}
\label{fig:3dscore}
\end{figure}

For training, we adopt the same procedure used in the synthetic case, with one minor modification: Rather than employing 2D patches, we utilize 3D input patches with dimensions of \(32 \times 32 \times 3\) (inline \(\times\) crossline \(\times\) time), corresponding to the physical size of 576 $\mathrm{m}$ $\times$ 576 $\mathrm{m}$ $\times$ 6 $\mathrm{ms}$. We select 1.7 to 1.9 sec for the monitoring area, which includes the reservoir area (1.8 to 1.9 sec) \cite[]{roach2017initial}. We use the image difference (PI2-PI1) for training. We train the autoencoder first by minimizing Equation \ref{eq:argmin} with the mean squared error (MSE) loss function (Equation \ref{eq:mse}). The learning rate is 0.0001, and the total number of epochs is 100. Then, using the trained encoder, we extract the 128-dimensional latent vector from the patches and determine the sphere's center (c) by averaging those vectors. The encoder is then fine-tuned by minimizing Equation \ref{eq:deep-svdd}. We employ a learning rate of 0.0001 and a total of 10 epochs for fine-tuning. Then, we predict the anomaly scores from both image differences: 'M1-PI1' and 'M2-PI1' (Figures \ref{fig:3dscore} (a,b)). Finally, we get the DeepNRMS differences for M1 and M2 (Figures \ref{fig:3dscore} (e,f)) by multiplying the predicted scores for corresponding NRMS differences (Figures \ref{fig:3dscore} (c,d)). As we can see in Figures \ref{fig:3dscore}, the DeepNRMS differences for both M1 (c) and M2 (d) show the plumes more clearly than the original NRMS differences. In the results, we observe the upward movement of the plume in M2 (red arrow in Figure \ref{fig:3dscore} (f)). In contrast, this was not observed in M1 (red arrow in Figure \ref{fig:3dscore} (c)), suggesting that this movement is likely due to the expansion of the plume by additional injection.

\begin{figure}[ht!]
\centerline{\includegraphics[width=1\textwidth]{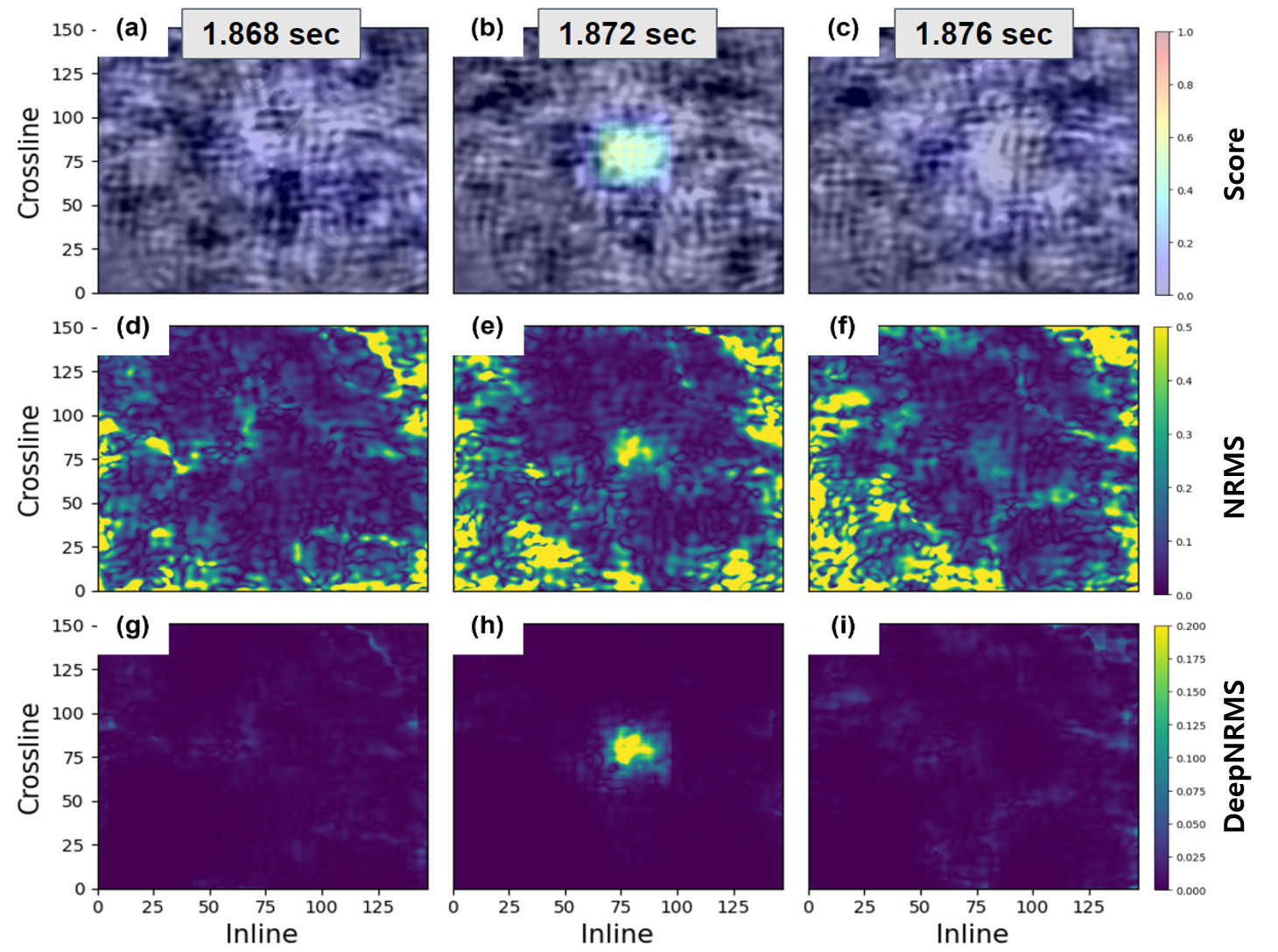}}
\caption{The time slices (a,b,c) of the amplitude differences (M1 and PI1) with the superposed anomaly scores. (d,e,f) The NRMS differences (M1 and PI1) and (g,h,i) the DeepNRMS differences (M1 and PI1). Each column represents the different times (left to right): 1.868 sec, 1.872 sec, and 1.876 sec. }
\label{fig:map_m4}
\end{figure}

\begin{figure}[ht!]
\centerline{\includegraphics[width=1\textwidth]{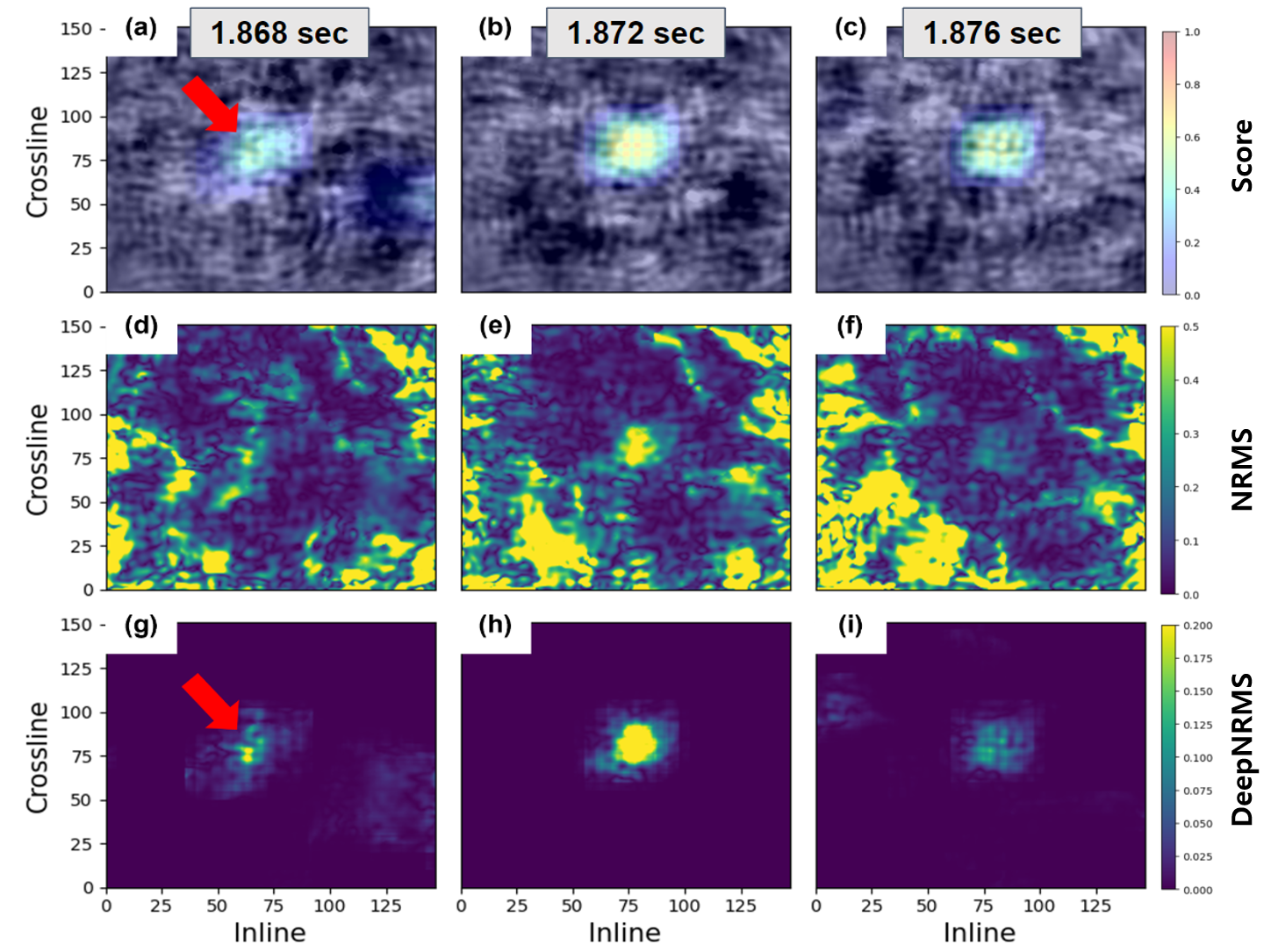}}
\caption{The figures' structures are the same as Figure \ref{fig:map_m4}, but for M2 and PI1.}
\label{fig:map_m5}
\end{figure}

Further analysis is performed through a time slice section to provide a more comprehensive understanding of the results. The top row of Figure \ref{fig:map_m4} shows the time slices of the amplitude differences between PI1 and M1. Note that the anomaly scores are superposed. The middle row shows the NRMS differences, and the bottom row shows the DeepNRMS differences. Each column represents the different times (left to right): $1.868$ sec, $1.872$ sec, and 1.876 sec. As we can see, the anomaly score successfully filters out the majority of high NRMS differences except the one caused by the plume (Figure \ref{fig:map_m4} (h)). Figures \ref{fig:map_m5} show the same results for M2. Note that the configuration of Figures \ref{fig:map_m5} is the same as Figures \ref{fig:map_m4}. Similar to M1 results, it removes most of the high NRMS differences at the edges. However, (red arrows) there are high anomaly scores at the shallower area (Figure \ref{fig:map_m5} (g)), which are not observed in M1. It implies that the plume has moved upward.

\begin{figure}
\centerline{\includegraphics[width=1\textwidth]{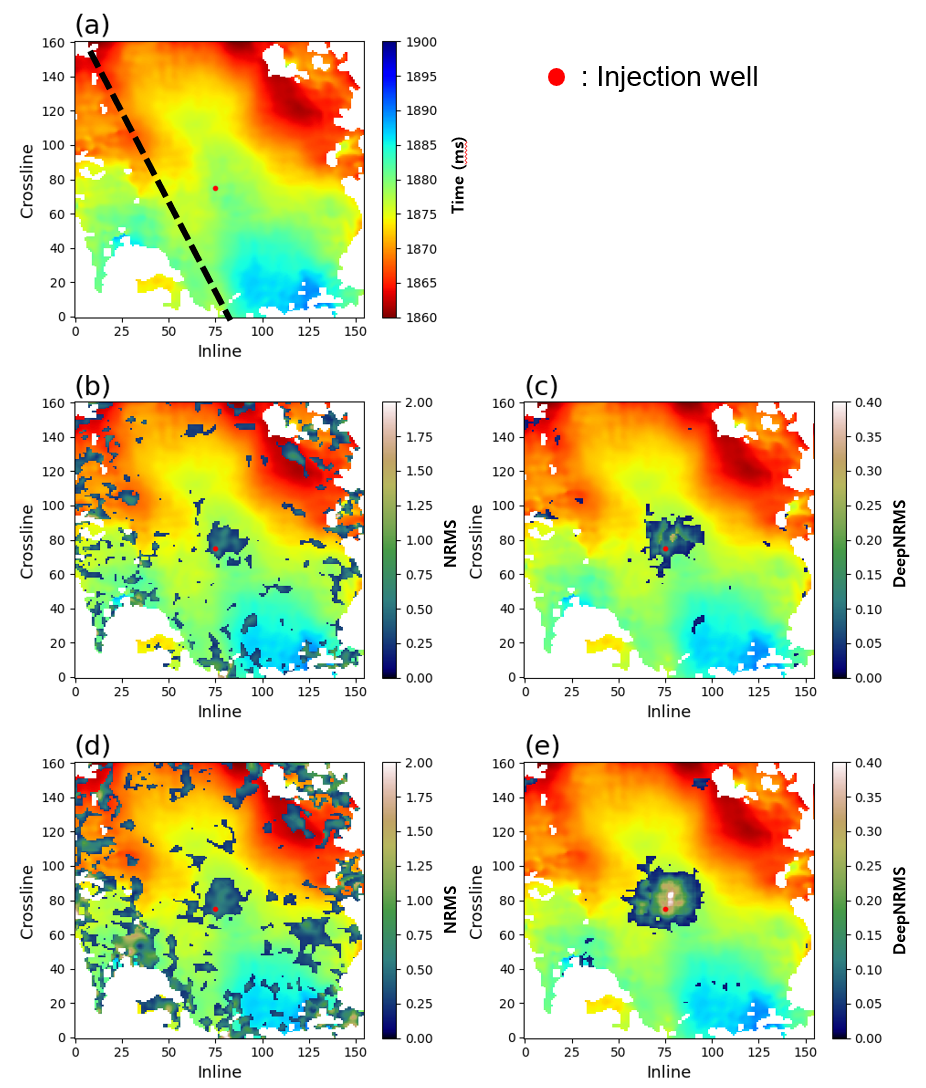}}
\caption{The picked horizon of the reservoir (a) overlaid with (b) M1 NRMS, (c) M1 DeepNRMS, (d) M2 NRMS, and (e) M2 DeepNRMS. The red dot indicates the injection well's location. Note that all NRMS and the DeepNRMS are taken from the corresponding horizon.}
\label{fig:horplume}
\end{figure}

To provide a comprehensive analysis of the plume behavior, we conducted horizon picking of the reservoir layer. Figures \ref{fig:horplume} show the picked horizon of the reservoir (a), overlaid with (b) M1 NRMS, (c) M1 DeepNRMS, (d) M2 NRMS, and (e) M2 DeepNRMS. The red dot indicates the injection well's location. Our findings confirm the presence of flexure structures along the black dashed lines in Figure \ref{fig:horplume} (a), which aligns with previous studies \cite[]{white2016geological,white20183d,roach2018evolution}. To enhance visualization, we applied a thresholding step, removing NRMS values below $0.2$ and DeepNRMS values below $0.02$. It is important to note that all NRMS and the DeepNRMS values were extracted from the corresponding time of the horizon. As depicted in Figures \ref{fig:horplume}, our proposed method effectively mitigates a significant portion of the time-lapse noise present in the original NRMS, particularly near the edges. Additionally, the DeepNRMS enhances the signals in the vicinity of the injection point. The results show consistent and continuous shapes, which instills confidence in the accuracy of our approach.

\begin{figure}
\centerline{\includegraphics[width=1\textwidth]{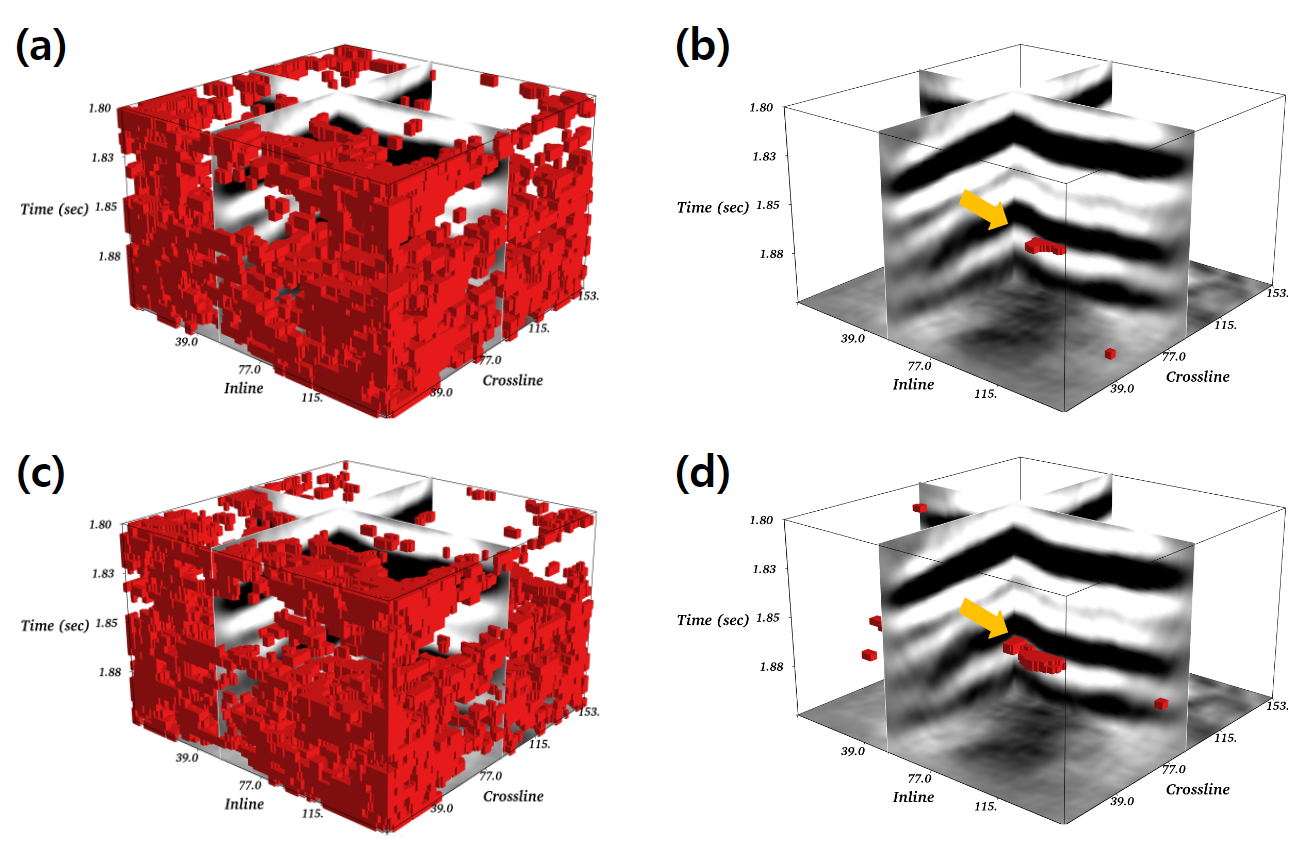}}
\caption{\ch{CO2} plumes locations (red) after applying threshold are shown on the PI1 images. The first column shows the plume prediction on the NRMS difference for (a) M1 and (c) M2, while the second column represents the DeepNRMS prediction for (b) M1 and (d) M2.}
\label{fig:3dplume}
\end{figure}

We also visualized the results by localizing the \ch{CO2} plumes on 3D seismic image cubes, as depicted in Figure \ref{fig:3dplume}. Initially, we applied thresholds to both the NRMS and the DeepNRMS differences to isolate the desired plume locations. Subsequently, we mapped these locations onto the 3D seismic image cubes. The first column of Figure \ref{fig:3dplume} showcases the plume predictions based on the NRMS difference for (a) M1 and (c) M2, while the second column presents the DeepNRMS predictions for (b) M1 and (d) M2. Our observations indicate that the predicted plume location obtained from the proposed method matches the layer in the seismic image. Furthermore, the majority of time-lapse noise has been effectively removed, enabling us to identify the \ch{CO2} plume signals more clearly. Particularly noteworthy is the DeepNRMS prediction for M2 (Figure \ref{fig:3dplume} (d)), which demonstrates an upward movement following the updip direction of the upper layer, as indicated by the yellow arrows. This phenomenon suggests that buoyancy forces likely play a significant role in driving the observed movement.
%--------------------------------------------------------------------%

%--------------------------------------------------------------------%
\section{Conclusions}
The proposed unsupervised deep learning-based \ch{CO2} monitoring system demonstrates great potential in accurately detecting subsurface changes in time-lapse seismic surveys, even in the presence of strong time-lapse noise. One of the notable advantages of our method is its unsupervised nature which can eliminate the need for labeled data and enables more efficient learning of time-lapse noise characteristics. While the proposed method necessitates supplementary pre-injection surveys, this constraint can be minimal, especially considering long-term monitoring. In both synthetic and field examples, our method successfully removes a significant portion of time-lapse noise and enhances the \ch{CO2} plume signal. While the results from both examples showcase the effectiveness of our method, it is important to acknowledge a few limitations. One limitation is when the trained model is confronted with new types of time-lapse noise that were not present in the training data. This is because our method does not specifically target \ch{CO2} plumes, and additional fine-tuning steps may be required to handle such situations effectively. Additionally, determining the optimal amount of pre-injection data required for each specific case remains a challenging task.

\section{Acknowledgement}
We express our profound gratitude to Don White and the Petroleum Technology Research Centre for generously providing the Aquistore seismic data. Our heartfelt thanks are also extended to the affiliates of the Stanford Earth Imaging Project for their invaluable financial support of this research. Additionally, we acknowledge Stanford University, Lawrence Livermore National Laboratory, and TotalEnergies for their contributions to the GEOS open-source simulator.
\newpage

\bibliographystyle{seg}  % style file is seg.bst
\bibliography{paper}

\end{document}